\documentclass[prodmode,acmtap]{ci2018}

\acmVolume{2}
\acmNumber{3}
\acmArticle{1}
\articleSeq{1}
\acmYear{2010}
\acmMonth{5}

\usepackage[ruled]{algorithm2e}
\usepackage[hyperindex,breaklinks]{hyperref}
\usepackage{bookmark}
\SetAlFnt{\algofont}
\SetAlCapFnt{\algofont}
\SetAlCapNameFnt{\algofont}
\SetAlCapHSkip{0pt}
\IncMargin{-\parindent}
\renewcommand{\algorithmcfname}{ALGORITHM}

\title{Crowd Work on a CV? Understanding How AMT Fits into Turkers' Career Goals and Professional Profiles}

\author{ANNA KASUNIC \affil{Human-Computer Interaction Institute, Carnegie Mellon University}
CHUN-WEI CHIANG \affil{Human Computer Interaction Lab, West Virginia University}
GEOFF KAUFMAN \affil{Human-Computer Interaction Institute, Carnegie Mellon University}
SAIPH SAVAGE
\affil{Human Computer Interaction Lab, West Virginia University}
}

\begin{abstract}
\end{abstract}

\begin{document}

\maketitle

\section{Introduction}
In 2013, scholars laid out a framework for a sustainable, ethical future of crowd work, recommending career ladders so that crowd work can lead to career advancement and more economic mobility \cite{Kittur:2013:FCW:2441776.2441923}. Five years later,  we consider this vision in the context of Amazon Mechanical Turk (AMT).

To  understand how workers currently view their experiences on AMT, and how they publicly present and share these experiences in their professional lives, we conducted a survey study with workers on AMT (n=98). The survey we administered included a combination of multiple choice, binary, and open-ended (short paragraph) items gauging Turkers' perceptions of their experiences on AMT within the context of their broader work experience and career goals. This work extends existing understandings of who crowd workers are and why they crowd work by seeking to better understand how crowd work factors into Turkers' professional profiles, and how we can subsequently better support crowd workers in their career advancement. Our survey results can inform the design of better tools to empower crowd workers in their professional development both inside and outside of AMT.  

\subsection{Relevant Literature}
This work builds upon existing understandings of who crowd workers are, e.g., primarily from the U.S., relatively highly educated and gender-balanced, with incomes lower than U.S. averages \cite{berg2015income}, \cite{difallah2018demographics}, \cite{ross2010crowdworkers}. Extant research lays out frameworks for understanding crowd workers' intrinsic (e.g., as a form of leisure; enjoyment of skill variety and task autonomy) and extrinsic (e.g., payment, gaining skills) motivations \cite{kaufmann2011more}, \cite{tokarchuk2012analyzing}, \cite{Antin:2012:SDB:2207676.2208699}, but do not explore how such motivations directly fit into their career trajectories. Moreover, we note that many studies use motivation frameworks with an end goal of learning how to better encourage workers to produce efficient, high-quality work for requesters \cite{Dow:2012:SCY:2145204.2145355}, \cite{kulkarni2012mobileworks}, rather than learning how to improve the user experience for Turkers. 

In our study, we view crowd workers as end users, and take a user experience approach to meeting their needs and desires. We draw on ethical concerns that crowd work is prone to low wages \cite{hara2017data}, \cite{deng2013crowdsourcing}, and that crowd work may alienate workers from employers and the fruits of their labors \cite{hansson2017alienation}. Work that has sought to shift power dynamics in existing platforms and recognizes agency and action among crowd workers inspires us and guides our ideation processes, e.g.  \cite{irani2015cultural,irani2016stories}. Further, as our analysis will show, certain groups of individuals may have unique struggles that affect their career goals and perceptions. Existing research has considered crowd work benefits and challenges for specific populations, such as older adults \cite{Brewer:2016:WAT:2858036.2858198}, people with disabilities \cite{Ding:2017:SEW:2998181.2998282,Zyskowski:2015:ACU:2675133.2675158}, and people with autism \cite{Hara:2017:IPA:3132525.3132544}, but our study exposes other ways in which crowd workers may be vulnerable and could benefit from targeted support to empower them in their professional lives. 

\section{Survey Design and Analysis}
We modeled many of our basic demographic and platform use questions on \cite{berg2015income} in order to allow comparison with other work, but will focus our analysis and discussion on the closed- and  open-ended questions we asked regarding workers' descriptions of the skills pertaining to their general career, and to their work on AMT, including whether they list AMT experience on their resum\'es or CVs; how they would convince a requester to hire them on AMT; and their current and past job titles and responsibilities. After iterating on and pilot-testing our (IRB-approved) survey, we ran a study on AMT (n=98). We initially compensated workers \$4 per HIT (completing the survey), but later gave everyone a bonus of \$8 after reviewing AMT recorded times and self-reported survey times to ensure fair pay; most Turkers spent about 30-45 minutes taking the survey. We analyzed the results qualitatively using open, iterative coding. (Coding schema and survey instrument available upon request).

\section{Findings}
In most respects, our 98 survey participants were similar in demographics to previous survey samples of crowd workers, with age ranging from 21-67 (at an average of 36), with relatively high education levels (56 had completed college or beyond). Where other samples had fairly even gender splits, our sample skewed male (62 male; 36 female), and was also more US-based (90/98). Average income was slightly lower than previous work, at a mean of \$36,692 and a median of \$30,000; 31 respondents said AMT was their primary source of income, but 87\% of the respondents held other jobs concurrently, and 75\% had prior work experience. Work experiences ranged widely, and included health care work, IT and computer programming, construction, law, manufacturing, landscaping, service industries, and finances; 12 were unemployed directly prior to starting crowd work. Participants held positions from entry level to advanced, with 23 individuals holding management positions either now or in the past.

When asked whether they currently list their experience on AMT on their resum\`e or CV, only 19 participants said yes, explaining that they wanted to show they had stayed busy during unemployment, that they felt they had gained valuable skills through AMT, and that it showcases certain skills such as versatility and ability to work independently. Yet the majority of respondents chose not to list AMT experience; while many of these simply hadn't updated or didn't need a resum\'e (n=28), 22 viewed the work as not relevant to their current career and job prospects, and 14 viewed their work on AMT as too trivial or unimpressive to list (not a ``real job''). As p50 wrote, \emph{``It just seems too small of a job to list on a resum\'e.''} Another 15 respondents feared that employers would not understand or recognize crowd work as valuable. One participant (p67) expressed, \emph{``I find it kind of hard to explain and embarrassing.''} 

Nonetheless, when asked to describe their skills both more broadly and on AMT, specifically, participants gave positive, actionable responses. Participants used may of the same terms to describe the skills they used on AMT and in their other jobs, though the distributions varied. For example, there were certain skills that tended to be more present in AMT work, such as attention to detail (16 for AMT; 5 for other jobs), speed (20 AMT; 3 other jobs), and honesty (20 AMT; 3 other jobs). Although relatively few respondents felt there was merit to adding AMT experience to their resum\'e, all but three wrote compelling self-promotions regarding why an AMT requester should hire them, citing their diligence, their dedication, and their approval ratings on the platform. These responses suggested that participants could see their value to AMT requesters, even if they struggled to find value in and/or communicate that value outside of AMT. Yet the dissenters to this question raised solid points regarding the expectations of the platform; unlike traditional employee-employer relationships, requesters on AMT do not formally ``hire" Turkers, so there is not the same need to impress or convince. As one respondent who said he would tell the requester to ``F*** off'' explained, \emph{``I'm not begging for anything. if you want me to do your hit and pay me for it, i'll do it. otherwise go find someone else'' [sic]} (p56).

 In discussing their future career goals, 24 expressed a desire to advance in their main career outside AMT. For example, ``I would to like to become a member of the Top management team member of my company'' (p53), and ``I would like to continue to expand my trading operation and either eventually start managing others' funds or increase my own funds to the point that I can make a real difference in getting projects off the ground'' (p80). A high proportion of these advancement-focused respondents (18/24) did not see AMT as contributing to their future goals. However, across the sample, 60/98 participants \textit{did} view AMT as contributing to their career goals. For example, of the 16 respondents who expressed desires for new career directions, 12 said they felt AMT is helping them approach these goals. Ten participants also expressed desires to improve their skills and earning ability on AMT. 

However, not all future career goals were quite as sunny. Fourteen responses indicated uncertainty about the future, at times to a troubling degree. Some of these responses were not so dire, but just indicated a general lack of direction, such as ``honestly not completely certain at this point'' (p10). Other responses raise more cause for concern, for example: ``Other people have future career goals for me, but I do not have my own. I don't believe that I can handle doing any more than I am already doing. I tried in the past and failed miserably'' (p35), and ``Not dying through suicide'' (p61). In considering the greater context, although we did not ask our participants specifically about any disabilities or mental health issues, three respondents discussed disabilities, one discussed poor physical health, and three discussed mental health challenges, including panic attacks and post-traumatic stress disorder.

\section{Discussion}
In this exploratory study, we set out to understand how we can better support crowd workers in meeting their future career goals, and in communicating the value of their AMT work to others, including potential employers. Although we initially had ideas in mind for systems to help automate and suggest ways to self-present, and identifying HITs that could help Turker gain skills pertinent to their career goals, our findings suggest there may be other avenues that should take precedence. For example, there may be subsets of users that struggle with mental health issues that may impede their abilities to formulate and move towards career goals. Although these users represented a small portion of our sample, we believe providing support for vulnerable populations that work on AMT is crucial, and that this group may still represent a large number of people when considering the entire population of AMT workers. Further investigation and intervention may also need to focus on exploring the attitudes and perceptions of (external) employers; it appears that Turkers can speak of their accomplishments and skills on AMT with ease, but that for a subset of users, getting employers to understand and accept AMT poses a challenge. Other users have uncertainty about the next steps in their career, and use AMT as a ``holdover'' until they come up with a better plan; general career guidance rather than skill-building guidance could prove useful to these Turkers. Lastly, there may be new ways to measure performance on AMT by leveraging the skills and attributes that Turkers themselves value, namely honesty; an honesty metric, gained over time, could prove useful both to Turkers and to requesters.
\subsection{Next Steps}
We plan to explore the design space using UX research methods such as visual improvization \cite{luma2012innovating} and speed dating, a method similar to sketching and prototyping for ``rapidly exploring application concepts and their interactions'' that helps designers identify critical issues and better understand users' needs \cite{davidoff2007rapidly}.  Sensitive to Irani and Silberman's critiques of designers posing as the agents of change \citeyear{irani2016stories}, we look forward to engaging directly with Turkers as co-designers.

\subsection{Acknowledgements}A special thanks to Kristy Milland for her feedback, to the NSF GRFP, and to all our participants.

\bibliographystyle{ci-format}
\bibliography{main}

\end{document}